# Hierarchical network design for nitrogen dioxide measurement in urban environments, part 2: network-based sensor calibration

Lena Weissert[1,3][a], Elaine Miles[2], Georgia Miskell[1b], Kyle Alberti[2], Brandon Feenstra[4], Geoff S Henshaw[2], Vasileios Papapostolou[4], Hamesh Patel[2], Andrea Polidori[4], Jennifer A Salmond[3], David E Williams[1,*].

*Email  david.williams@auckland.ac.nz    ph +64 9 923 9877

1. School of Chemical Sciences and MacDiarmid Institute for Advanced Materials and Nanotechnology, University of Auckland, Private Bag 92019, Auckland 1142, New Zealand
2. Aeroqual Ltd, 460 Rosebank Road, Avondale, Auckland 1026, New Zealand
3. School of Environment, University of Auckland, Private Bag 92019, Auckland 1142, New Zealand
4. South Coast Air Quality Management District, 21865 Copley Drive, Diamond Bar, CA 91765, USA

**Abstract**

We present a management and data correction framework for low-cost electrochemical sensors for nitrogen dioxide ($NO_2$) deployed within a hierarchical network of low-cost and regulatory-grade instruments. The framework is founded on the idea that it is possible in a suitably configured network to identify a source of reliable 'proxy' data for each sensor site that has a similar probability distribution of measurement values over a suitable time period. Previous work successfully applied these ideas to a sensor system with a simple linear 2-parameter (slope and offset) response, with parameters estimated by moment matching site and proxy data distributions. However, applying these ideas to electrochemical sensors for $NO_2$ presents significant additional difficulties for which we demonstrate solutions. The three $NO_2$ sensor response parameters (offset, ozone ($O_3$) response slope, and $NO_2$ response slope) are known to vary significantly as a consequence of ambient humidity and temperature variations. Here we demonstrate that these response parameters can be estimated by minimising the Kullback-Leibler divergence between sensor-estimated and proxy $NO_2$ distributions over a 3-day window. We then estimate an additional offset term by using co-location data. This offset term is dependent on climate and spatially correlated and can thus be projected across the network. Co-location data also estimates the time-, space- and concentration-dependent error distribution between sensors and regulatory-grade instruments. Robust $O_3$ measurements are obtained using a semiconducting oxide-based instrument, previously described. We show how the parameter variations can be used to indicate both sensor failure and failure of the proxy assumption. With these procedures, we demonstrate measurement at nine different locations across two regions of Southern California over seven months with average root mean square error ± 7.2 ppb (range over locations 4 – 11 ppb) without calibration other than the remote proxy comparison. We apply the procedures to a network of 56 sensors distributed across the Inland Empire and Los Angeles County regions. The results show large variations in $NO_2$ concentration taking place on short time- and distance scales across the region. These spatiotemporal $NO_2$ variations were not captured by the more sparsely distributed regulatory network of air monitoring stations demonstrating the need for reliable data from dense networks of monitors to supplement the existing regulatory networks.

---

[a] Present address:  Aeroqual Ltd, 460 Rosebank Road, Avondale, Auckland 1026, New Zealand
[b] Present address: Trustpower, 108 Durham St, Tauranga, New Zealand

## 1. Introduction

The question of reliability of data from low-cost sensors is contentious and difficult to address (Williams, 2019). An approach that uses independent information to support sensor data is promising. We present one such approach here, applied to measurement of nitrogen dioxide with electrochemical cells, that extends previously described methods for $O_3$ (Miskell et al., 2016; Miskell et al., 2018). Advancement in technology has resulted in the availability of low-cost sensors that can be used to collect real-time $NO_2$ data at a high spatial and temporal resolution (Snyder et al., 2013). When deployed in dense hierarchal networks, low-cost sensors offer an opportunity to collect neighbourhood-level air pollution data. They have been used to detect small scale variations (Mead et al., 2013) and discriminate emissions due to different activities and emission sources (Popoola et al., 2018). Thus, they have become a popular choice for community-based air quality networks and community science projects (Clements et al., 2017; Hubbell et al., 2018). However, uncertainties remain about the data reliability of low-cost $NO_2$ sensors largely due to drift and interferences with other pollutant gases and variations associated with changes in temperature and relative humidity (Isiugo et al., 2018; Lewis et al., 2016; Mead et al., 2013; Weissert et al., 2019). In an attempt to calibrate the sensors and assess their accuracy, sensors are typically co-located against a well-maintained regulatory reference instrument for a period of time before and after deploying them in the field (Isiugo et al., 2018; Sadighi et al., 2018; Weissert et al., 2019). This appears a suitable approach only for short term deployments, while long-term deployments would require ongoing re-calibration (van Zoest et al., 2019) leading to calibration and maintenance costs that may quickly exceed the costs of the instruments (Clements et al., 2017). In addition, this approach assumes that the calibration parameters obtained from the co-location of the low-cost sensors at a reference site are transferable to other locations in the sensor network. A recent study from a network of $NO_2$ sensors in Eindhoven, Netherlands has shown that the calibration coefficients could not easily be transferred from one location to another within a city likely due to drift and interference effects being different for individual sensors (van Zoest et al., 2019) and to significant time-variation of the individual sensor response parameters. One suggestion to overcome this problem is the use of a mobile reference sensor that is moved from one location to another for calibration, which would account for the spatial and temporal differences in the calibration parameters (van Zoest et al., 2019). However, the costs associated with this approach may quickly outweigh the benefits of the low-cost sensors particularly if they are deployed in dense networks.

In our previous work, we developed a semi-blind management framework to verify the reliability of low-cost sensor data using general knowledge of the sensor and pollutant. Consequently, we were able to demonstrate remote correction of low-cost sensors that are deployed in dense networks (Alavi-Shoshtari et al., 2013; Miskell et al., 2016; Miskell et al., 2018, Miskell et al. 2019). The management framework was tested using hierarchical networks, consisting of well-maintained regulatory-grade instruments and low-cost $O_3$ sensors deployed around the Lower Fraser Valley (LFV) in Canada (Miskell et al., 2018) and Southern California (Miskell et al., 2019). Data from the well-maintained regulatory-grade instruments were first used to determine suitable proxies across the region, and then to provide suitable proxy data to check for drift and if necessary apply a correction. We defined a proxy as a reliable source of data within the network but at a different location to the site of interest, whose data has a similar probability distribution (Miskell et al., 2018). A proxy site can be selected based on proximity or similar land use (Miskell et al., 2019; Miskell et al., 2016). Testing this approach in these two distinct regions, which differ considerably in terms of geography, traffic patterns, climate and population density, suggests that the approach is transferable. In a previous paper, we tested the possibility of selecting a suitable proxy for $NO_2$ using regulatory data, which are frequently and rigorously calibrated (Weissert et al., 2019b, submitted). The results showed that even for pollutants like $NO_2$, which is highly variable spatially and temporally, a suitable proxy can be selected.

The purpose of this paper is to extend the management framework to electrochemical sensors for $NO_2$, where the measurement model for the sensor is more complex than a simple 2-parameter model, and where interfering effects of climate variables are also complex.

## 2. Methods

*2.1 Study sites*

The study sites were distributed across the Los Angeles region (Figure 1). There are five regulatory sites in the Los Angeles city ('LA') and four sites in the Inland Empire ('IE') which includes Riverside and San Bernardino Counties in Southern California (Figure 1). The sites are equipped with continuous reference method Nitrogen Oxides (NOx) analyzers, which are regularly maintained and serviced by the South Coast Air Quality Management District (South Coast AQMD). Eight sites are equipped with a model 42i NOx analyzer by Thermo Fischer Scientific (Franklin, MA), while the Fontana site is equipped with a model 200E NOx analyzer by Teledyne Advanced Pollution Instrumentation (San Diego, CA). At each site, we had a low-cost instrument that measures $O_3$ and $NO_2$ (details below: AQY1, Aeroqual, Auckland, NZ). We used data from January – July 2018 for the IE network and from March – July for the LA network. Vehicle emissions, particularly from heavy-duty vehicles, are the main source of $NO_2$ in the LA region (AQMP, 2016). Nitrogen oxides (NOx) are precursors to both $O_3$ and particulate matter (PM) and therefore of major concern for air quality management (AQMP, 2016). Measurements are mixing ratios: parts-per-billion ($10^9$) by volume (ppb).

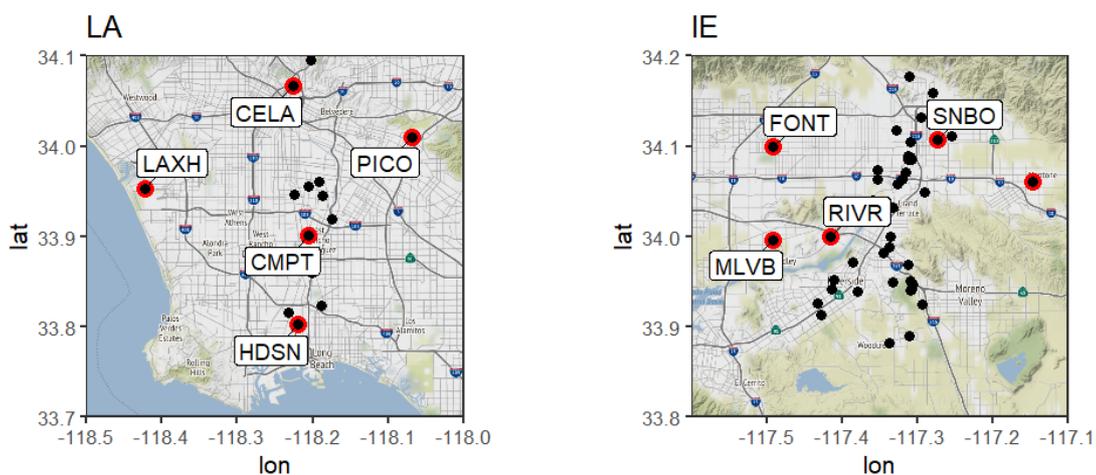

Figure 1. Map of the regulatory sites (red points) and the low-cost instruments (black points) in the Los Angeles (LA) and Inland Empire (IE) region. A low-cost instrument was co-located at each regulatory site where both $O_3$ and $NO_2$ are measured.

*2.2 Low-cost sensors*

The low-cost sensors deployed in the Los Angeles network are the AQY v0.5 sensors from Aeroqual Ltd, Auckland, New Zealand. We use the term 'sensor' here to refer both to the instrument package ($O_3$, $NO_2$, T, RH and $PM_{2.5}$) and to the detection element. $O_3$ was measured using a gas-sensitive semiconducting (GSS) oxide, $WO_3$, as the detection element (Aliwell et al., 2001; Hansford et al., 2005; Utembe et al., 2006; Williams et al., 2002). Air flow-rate modulation and temperature modulation are used to cancel interferences due to water vapour, and to continually reset and re-zero the sensor. This device has been shown to be robust, reliable and accurate for ambient monitoring (Bart et al., 2014; Miskell et al., 2018; Williams et al., 2013). $NO_2$ was measured using an electrochemical sensor, whose response has been characterised in detail (Weissert et al., 2019). $O_3$ and $NO_2$ measurements were collected with 1 min time resolution and then hourly-averaged. The instrument has been described in detail in Weissert et al. (2019a). The electrochemical $NO_2$ sensor element was supplied by Membrapor.

*2.3 Proxy selection*

The proxy sites for the $O_3$ and $NO_2$ correction were established using data from the well-maintained South Coast AQMD regulatory network deployed in the LA and Inland Empire region (Fig. 1) with the procedures described in Miskell et al., 2019, Miskell et al., 2016 and Miskell et al., 2018. The most suitable proxies for $O_3$ were sites in closest proximity to deployed sensors. For the $NO_2$ correction using the proxy with similar land use proved to be more suitable than the nearest site, with the exception of two regulatory sites in Mira Loma (MLVB) and Rubidoux (RIVR) located in a semi-closed valley (Fig. 1), for which the closest site was more appropriate (Weissert et al., 2019b submitted). The land use variables were chosen based on the most commonly used variables in published $NO_2$ land use regression (LUR) studies in the North American Region. The procedure for proxy choice has been discussed in detail in part 1 (Weissert et al., 2019b submitted). The proxy sites are described in Table 1 below.

Table 1. Selected proxies for the framework correction. $O_3$ was selected based on the nearest site and $NO_2$ was selected based on the most similar land use.

| AQY site | Regulatory site | $NO_2$ proxy site | $O_3$ proxy site |
| --- | --- | --- | --- |
| 100 | RIVR | MLVB | MLVB |
| 101 | MLVB | RIVR | RIVR |
| 102 | SNBO | MLVB | RIVR |
| 103 | FONT | SNBO | MLVB |
| 161 | PICO | CELA | CELA |
| 166 | CMPT | HDSN | HDSN |
| 176 | LAXH | CMPT | CMPT |
| 177 | HDSN | CMPT | CMPT |
| 182 | CELA | HDSN | PICO |

*2.4 Management framework*

The drift-detection framework is described in detail by Miskell et al. (2016). It is based on three different tests comparing the distribution of the measurement result, *Y*, with that of a proxy, *Z*, evaluated running over a time $t_d$ : the Kolmogorov-Smirnov test for significant difference between the distributions $Y$ and $Z$, $p_{KS} < 0.05$; and the estimates from moment-matching of apparent slope, $0.7 < \hat{a}_1 < 1.3$ and offset, $-5$ ppb $< \hat{a}_0 < 5$ ppb, where $\hat{a}_1 = \sqrt{\mathrm{var}[Z]/\mathrm{var}[Y]}$ and $\hat{a}_0 = \mathrm{E}[Z] - \hat{a}_1 \mathrm{E}[Y]$, with var[] denoting the variance and E[] the mean evaluated over $t_d$. When any of these conditions are not met for a duration of five days ($t_f$), an alarm is triggered and is used to indicate potential sensor drift. The probability distributions are determined over a window of three days, $t_d$. Data is corrected if one or more alarms are triggered.

The measurement model for the $NO_2$ sensor, relating the measured current in the electrochemical cell, $i_{meas}$, do the indicated concentration of $NO_2$, $C_{NO_2}$, is (Weissert et al. 2019a):

$$C_{NO_2} = b'_0 - b'_1 i_{meas} - b_2 C_{O_3} \qquad (1)$$

Factory calibration of the assembled instrument before field deployment determines a number, $C_{ox} = b'_1 i_{meas} - b'_0$, which is linearly related to the raw current measurement. The instrument reports $C_{ox}$ as well as the $NO_2$ concentration derived from the factory calibration and the uncorrected $O_3$ concentration determined with the $O_3$ sensor. Now, the offset, $b'_0$, and the response slopes, $b'_1$ and $b_2$, can be time-varying, for example in response to changes in atmospheric humidity or temperature. The objective of the procedure is to estimate and correct for this variation. The measurement model to be used, therefore, given the results reported by the instrument, is written

$$\hat{C}_{NO_2} = \hat{b}_0 + \hat{b}_1 C_{ox} - \hat{b}_2 C_{O_3} + e \qquad (2)$$

where $e$ denotes any signal not accounted for by the principal variables assumed to drive the response and which also includes any measurement noise. Following the concepts described earlier, the correction method estimates values of the parameters $b_j$ to match the probability distribution over time $t_d$ of the estimate $\hat{C}_{NO_2}$ to that of a proxy, $Z_{NO_2}$, by minimising a suitably chosen objective function. The proxy site is chosen based on land use similarity. We explored two methods which gave similar results. First, we evaluated minimisation of the sum of squared differences of the first three moments of the distributions. Second, we evaluated minimisation of the Kullback-Leibler divergence (see supporting information, SI, for definition) of the two distributions, $D_{KL}\big(\mathbb{P}(\hat{C}_{NO_2})||\mathbb{P}(Z_{NO_2})\big)$. The moment matching method emphasises the tails of the distributions. The Kullback-Leibler method, on the other hand, emphasises the most probable values, and its minimisation is equivalent to maximising the mutual information or minimising the relative information entropy of the two distributions. In the following sections, we present the results from the minimisation of $D_{KL}$. Thus, we aim to find best estimates $\hat{b}_j$ such that:

$$D_{KL}\left(\mathbb{P}(\hat{C}_{NO_2}|C_{ox}, C_{O_3}, \hat{b}_j)||\mathbb{P}(Z_{NO_2})\right) = min \qquad (3)$$

The distributions are obtained by computing histograms with fixed bin size. In this calculation, the value of $C_{O_3}$ used is that delivered by the $O_3$ sensor which is checked and corrected if necessary according to the management framework as previously described (Miskell et al., 2019). The parameters are re-estimated only when the comparison of the (previously) estimated $\hat{C}_{NO_2}$ with the proxy gives an alarm, thus minimising the computational overhead. The process is initiated using the concentration values given by the pre-deployment factory calibration, denoted here $C_{NO_2}$(raw). The probability distribution of the estimate should be a sum of three distributions corresponding to the three terms. The variability of $C_{ox}$ would be determined by the noise in the electrochemical sensor (Weissert et al., 2019) and the averaging approach used to reduce this. $O_3$ and $NO_2$ measurements were collected with 1 min time resolution and then were hourly-averaged. Based on the results reported previously, we expect the standard deviation of this number to be less than 1 ppb. The RMSE of $C_{O_3}$, corrected according to the management framework, is 5.4 ppb for all reference sites combined and the entire study period (January – August), with a maximum RMSE of 7 ppb for individual sites (Miskell et al., 2019).

Two issues could affect the reliability of the parameters in equation (2) obtained through minimisation of the difference between the probability distributions. First, if the distributions approximate simple 2-parameter distributions (e.g. log-normal) then deriving three parameters from the comparison over-fits the data and would raise issues of correlation between the parameter estimates. Figure 2 shows reference station data from both summer and winter, compared to a 2-parameter log-normal model. Some sites over the two seasons do approximate a simple log-normal model, which would cause issues with the method. However, for most locations, this does not apply: the site data distribution is significantly skewed to low values. Hence, in general, we do not expect an overfitting issue. Second, under circumstances where $O_3$ and $NO_2$ reported similar concentration levels, an unconstrained minimisation could easily lead to physically unreasonable estimates with the parameters changing sign.

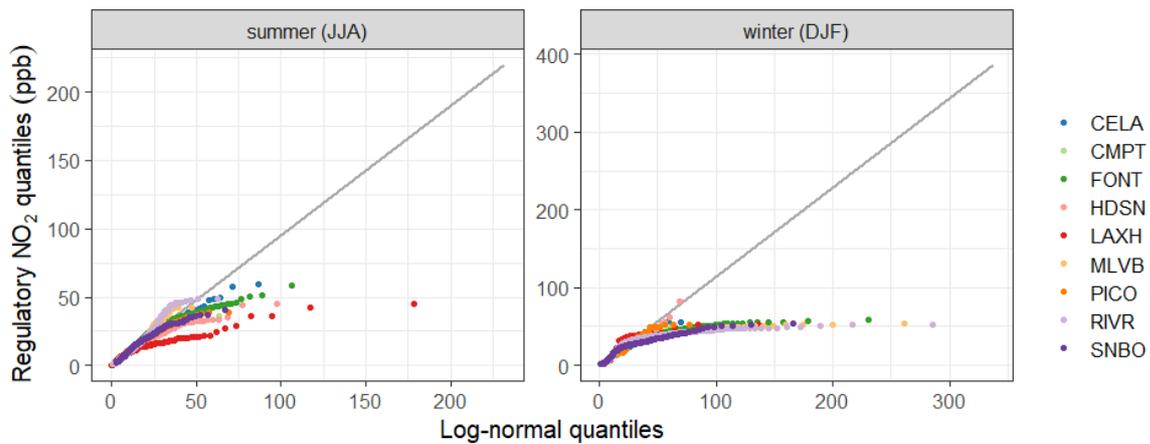

Figure 2. QQ-plots assessing the fit of reference station data to a log-normal distribution, for winter (January/February) and b) summer (June/July).

Indeed, we noted that minimisation of $D_{KL}$ with $\hat{C}_{NO_2}$ calculated with eq (2) without physically realistic initial estimates of the parameters, could easily lead to false minima with physically unrealistic parameter values (e.g. inverted sign). Physically realistic initial estimates for the minimisation were obtained as follows:

a) the measurement model is approximated by setting $b_2 = b_1$ as observed and also theoretically expected for an electrochemical sensor of this type without $O_3$ decomposition catalyst applied (Weissert et al., 2019).

b) the initial estimates of $b_0$ and $b_1$ (=$b_2$) are obtained by moment matching to the proxy:

$$\hat{b}_{2,init} = \hat{b}_{1,init} = \sqrt{\text{var}\langle Z_{NO_2}\rangle / \text{var}\langle C_{O_x} - C_{O_3}\rangle} \qquad (4)$$

$$\hat{b}_{0,init} = E\langle Z_{NO_2}\rangle - E\langle C_{O_x} - C_{O_3}\rangle \qquad (5)$$

following which the $b_j$ are iterated in eq (2) to minimise $D_{KL}$ (eq 3). The value $C_{ox}$ is the raw signal from the electrochemical sensor using the internal offset and slope values as above, hourly averaged, and (as noted above) the value of $C_{O_3}$ used is that delivered by the $O_3$ sensor, hourly averaged, checked and corrected if necessary according to the management framework as previously described (Miskell et al., 2019). The management framework is schematically illustrated in figure 3

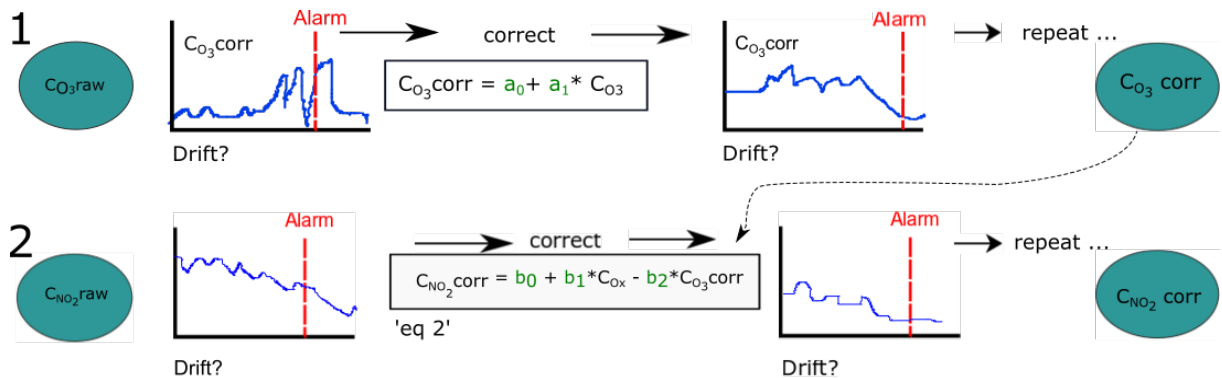

Figure 3. Summary of the $O_3$ and $NO_2$ management framework and correction process.

## 3. Results and Discussion

In this section, we first show the application of the Kullback-Leibler distribution matching method to co-location data, using the reference $O_3$ and $NO_2$ data from the co-location site. This application identified an offset error term that varied on a timescale less than the framework error detection timescale and therefore preventing the framework to compensate. The offset error was climate-related (mostly but not entirely ambient temperature) and spatially correlated. We used the knowledge that this offset error term was spatially correlated to apply an additional correction, derived using the closest proximity proxy site. Next, we applied the framework to sensors that were co-located at reference sites, but using proxy data and the sensor ozone data. By comparison with the reference data from the site of co-location, we were able to evaluate the error in the proxy procedure, using sensor ozone data.

*3.1 Using co-location data to evaluate the Kullback-Leibler method, sensor parameter variation and error terms*

Most co-location studies use regression methods. In contrast, our proxy comparison is based on similarity of probability distributions over a time interval. Therefore we used comparison of probability distributions on the co-location data to evaluate the performance of this method. For this part of the work we used the co-located reference $O_3$ data to avoid noise associated with the sensor $O_3$ correction. Figure 4 shows hexbin scatter plots of the sensor $NO_2$ against the co-location reference $NO_2$ over the 7 months of the study. The derived sensor parameter variations over time are given in figure S1. Parameter variation over time, within bounds, is expected. However, the sensors at MLVB and RIVR showed a downward drift of the slope parameters from July onward, very marked at MLVB, which the method compensated by an increase in the offset parameter. This behaviour should be taken as an indicator of sensor failure. All other sensors appeared stable. The hexbin plots show a significant scatter of the results. However, figure 5 shows that the difference between sensor-indicated $NO_2$ and reference $NO_2$ had a part that showed a diurnal variation as well as a part that showed apparently random variation. Figure 5 shows that the difference term was spatially correlated: the variations became larger at inland locations compared to those close to the sea. The correlation matrix is given in table S1 and the correlations between sites in closest proximity are shown in figure S2. The dependence of electrochemical $NO_2$ sensor signal on temperature, humidity and their rapid changes is known, but there is no simple relationship. Figure 5 also shows the joint probability distribution of the difference term and ambient temperature, measured by the sensor. Extreme values of the difference are associated with high (~50°C) or low temperature (< 12°C), but otherwise there is not a strong correlation.

The difference term can be attributed to large variations of the offset, $\hat{b}_0$. The variations are related to the electrochemical sensor, indicating fluctuations with a time scale between 1 hr and 3 days, these being the timescales of averaging and of comparison with the proxy distribution. Without a definitive model for the variations, it is difficult to provide a rigorously-based correction method. Below, we present an empirical method based on the observed spatial correlation.

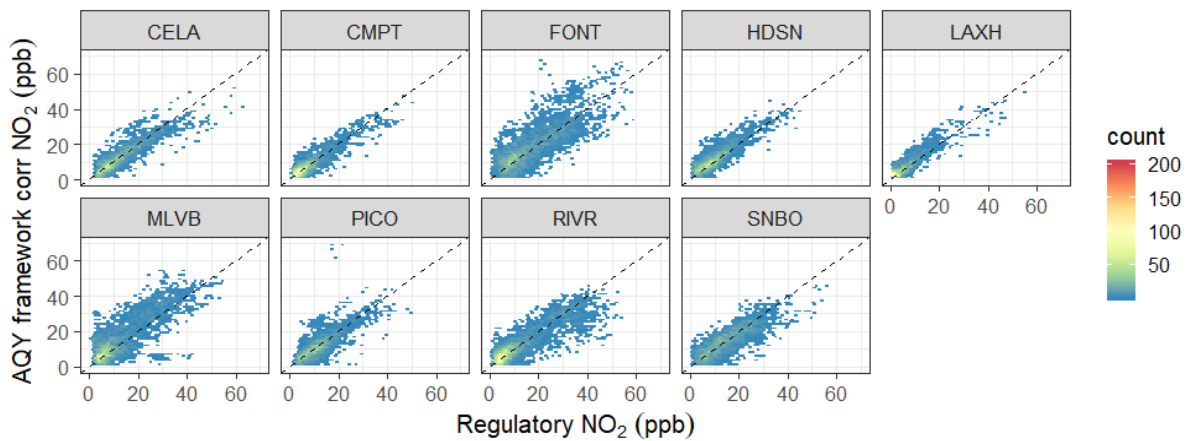

Figure 4. Hexbin scatter plots showing the correlation of sensor $NO_2$ with co-location reference $NO_2$, where the sensor $NO_2$ is derived using the distribution matching method (fig 3 and eq 2-5) and the co-location reference $NO_2$ and $O_3$ data.

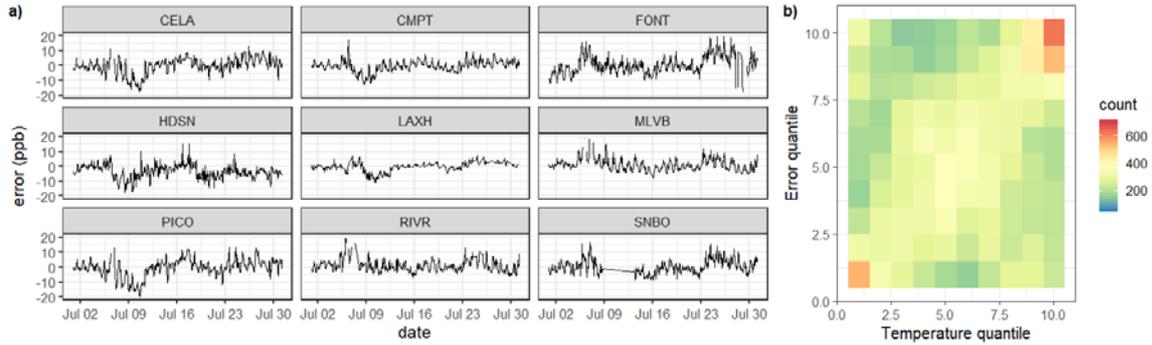

Figure 5. (a) Time series of the difference term between framework-derived sensor result and the co-location reference result. The site locations are in figure 1. (b) Joint probability distribution of difference term ('error') and ambient temperature, for the whole data set. Temperature Quantiles (°C): 0: 1, 1: 12, 2: 14, 3: 16, 4: 18, 5:19, 6: 21,7: 23, 8: 26, 9: 30, 10: 50.

Given these results, we rewrite the measurement model (eq 2) as:

$$\hat{C}_{NO_2} = \hat{b}_0 + \hat{b}_1 C_{ox} - \hat{b}_2 C_{O_3} + e_S + \varepsilon \qquad (6)$$

where $e_S$ denotes a spatially correlated error term and $\varepsilon$ the residual. Now, we propose a proxy method for evaluating $e_S$. Since we have electrochemical sensors co-located at reference sites, and the term is spatially correlated, an estimate of $e_S$ at some other site would be that value determined at the closest proximity reference site, at the required time. Figure 6 illustrates the issues with this idea. Firstly, if proxy data are unavailable at any particular period, then obviously no correction can be made; secondly, although the error term is spatially correlated on average, at any particular time, the difference between the values at the measurement site and the proxy site can be large. Given these issues, we found that this method could compensate a useful fraction of the difference term, provided the correction was limited: we used a sigmoid function to damp the error correction and a rolling average to smooth fluctuations; details are in the SI. Figure 7 shows hexbin scatter plots for the co-location data where the error term $e_S$ has been estimated from the closest proximity other site. The scatter is diminished at most sites. The overall RMSE improved and is 5 ppb (RMSE for individual sites in Table S2). Given that the estimated error due to sensor noise is less than 1 ppb, the major contributor to this error would be uncompensated sensor responses, such as are reflected in the uncompensated offset error term shown in figure 6.

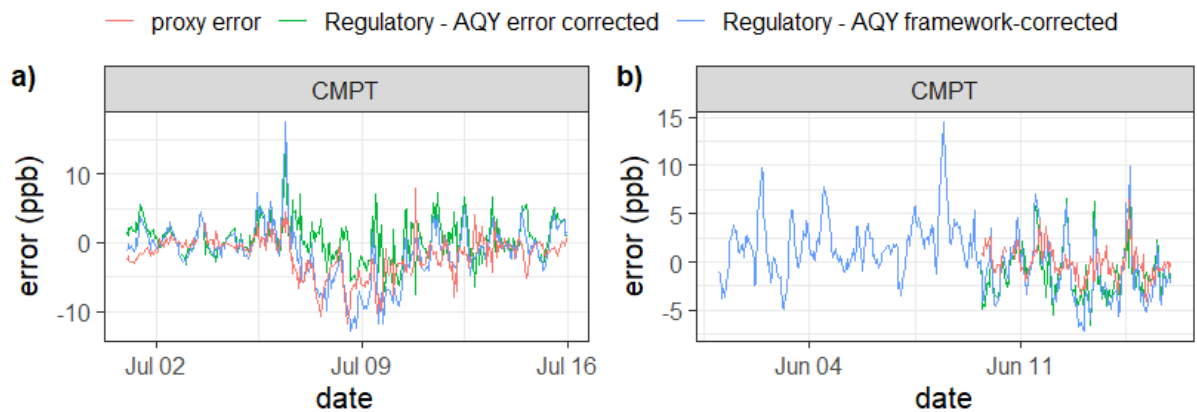

Figure 6. Examples of the uncompensated error term, and its partial correction using the error determined at the closest proximity site. red: error term determined at the closest proximity site; blue: actual error ($e_S+\varepsilon$, eq 6) determined at the measurement site following correction of the sensor using K-L method with the proxy site for $NO_2$; green: actual error determined at the measurement site following correction of the sensor using K-L method with the proxy site for $NO_2$ and determination of $e_S$ using the closest proximity site (damped and smoothed as described in the SI).

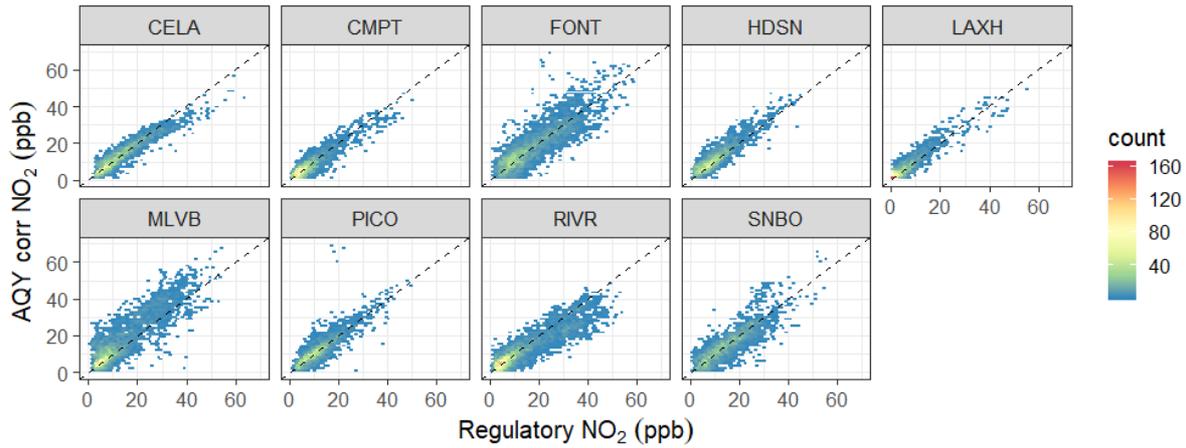

figure 7. Hexbin scatter plots showing the correlation of sensor $NO_2$ with co-location reference $NO_2$, where the sensor $NO_2$ is derived first using the framework distribution matching method (fig 3 and eq 2-5) and the co-location reference $NO_2$ and $O_3$ data, then by correcting using the closest proximity other co-location reference site to estimate $e_S$ (eq 6).

*3.2 Sensors at reference sites, using sensor ozone data and proxy sites to check and correct; assessment against co-location reference data*

Here, the framework was applied to sensors located at reference sites, using the $O_3$ sensor, the proxy sites for $NO_2$ (land use) and $O_3$ (proximity) to derive the sensor parameters using the K-L method according to eq 2-5, and the closest proximity proxy also to determine the spatially-correlated error, $e_S$ (eq 6).

Overall, the framework produced satisfactory results. Figure 8 shows examples of the time variation of the uncorrected and corrected rolling mean absolute bias (MAB) in relation to the co-located regulatory $NO_2$ and the alarm signals triggered over time at the regulatory sites. Data for all sites is in figure S5. Figure 9 shows the monthly average MAB at the different sites for the framework-corrected data. The management framework was able to detect and correct the drift resulting in a MAB within 2 and 10 ppb at most times and sites, which was a clear improvement to the uncorrected NO$_2$ MAB (up to 20 ppb) and considered satisfactory for an indicative air quality measurement (Snyder et al. 2013).

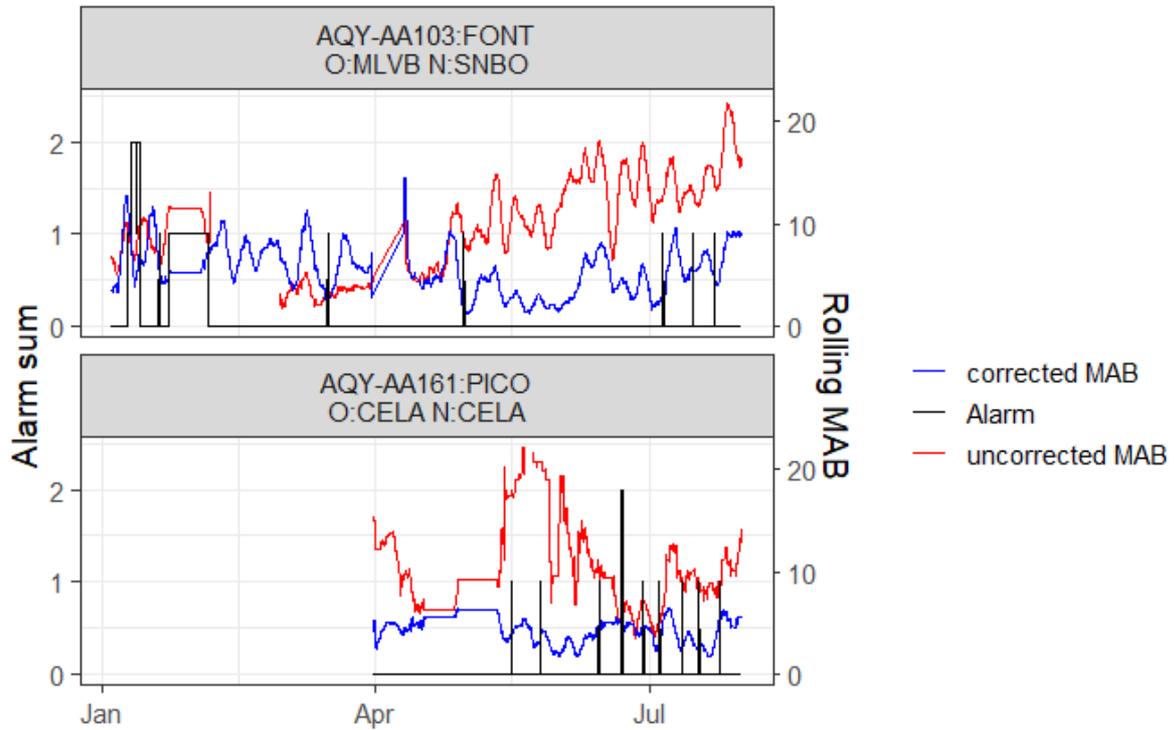

Figure 8. Illustrative examples showing the number of alarm signals generated by the low-cost sensor data in comparison with the proxy data (left-axis), and uncorrected vs. corrected mean absolute bias (MAB) running over 72 hr of the low-cost sensor data with respect to the co-located regulatory station (right-axis). One sensor from each region; top: FONT with O$_3$ and $e_S$ proxy MLVB and NO$_2$ proxy SNBO; bottom: PICO with O$_3$ and $e_S$ proxy CELA and NO$_2$ proxy CELA (shown in the label at the top of each chart).

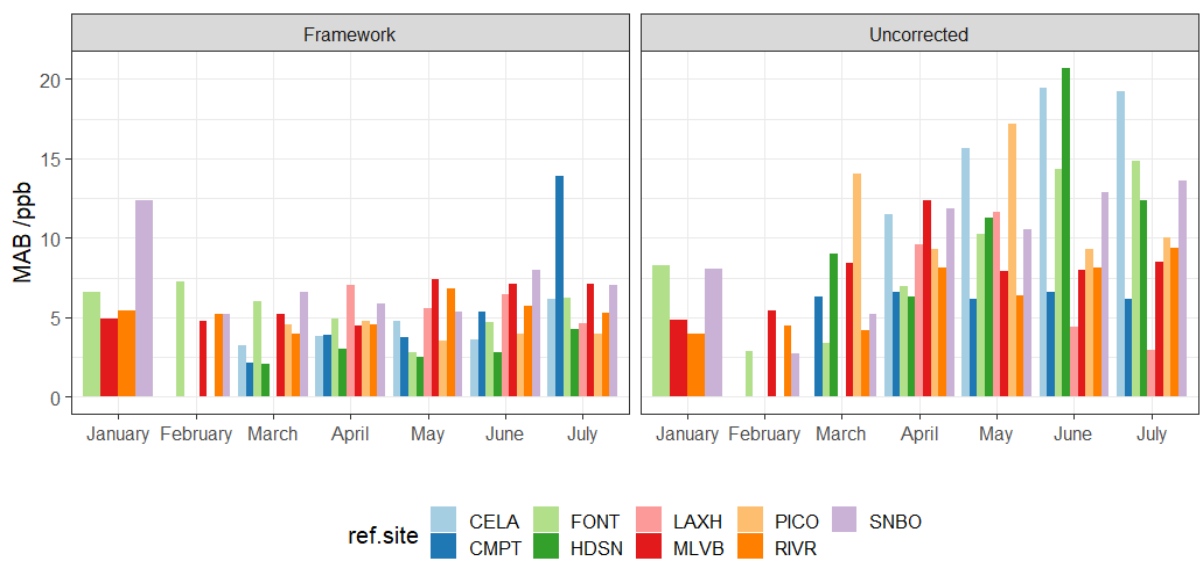

Figure 9. Mean Absolute Bias (MAB) compared to uncorrected MAB per month for proxy-corrected data ('Framework') at the different sites.

There was no obvious issue of over-fitting, which was consistent with the distributions being sufficiently different from a simple 2-parameter distribution. To illustrate the operation of the framework, figure 10 shows three examples: a site where the correction was satisfactory, though with a slight slope error (FONT); one where the sensor failed (MLVB); and one where the proxy selection was inappropriate for a particular time (CMPT, July: see also figure 9). Full data for all sites are given in figure S4-S6. For the sensor at FONT (correction and sensor satisfactory), figure 10 shows that the major variation over time was in the offset: indeed reflected in the MAB of the uncorrected data shown in figure 9. The two slope parameters were essentially constant and close to unity, with small fluctuations. The site and proxy distributions could be made almost coincident with a small alteration of response slope. The procedure resulted in the sensor data distribution being essentially coincident with the proxy distribution (evaluated over a month). For the sensor at MLVB, figure 9 shows a sudden and large jump in the offset during April which was associated with the start of a steady decrease in both the slope parameters. Inspection of the data showed the daily signal variations gradually decreasing towards zero. Although the correction in fact operated reasonably, clearly the sensor was failing, and there was an unknown event in mid-April that resulted in sensor failure. Monotonic change over time of the sensor parameters could be taken as indicative of sensor failure. For the sensor at CMPT, comparison of the data distributions given in figure 10 shows that, in July, the proxy and reference site distributions were very different. The procedure caused a bias in the sensor result towards the proxy with consequent over-estimation of the concentration. The response slope parameters both rose to values significantly greater than unity while the offset remained close to zero. The iterated minimum value of the objective function, $D_{KL}$, between corrected sensor data and the proxy became significantly larger. A strong variation of both slope parameters without corresponding variation of the offset, together with an increase in the iterated minimum value of the objective function could be assumed as indicative of an issue with the proxy.

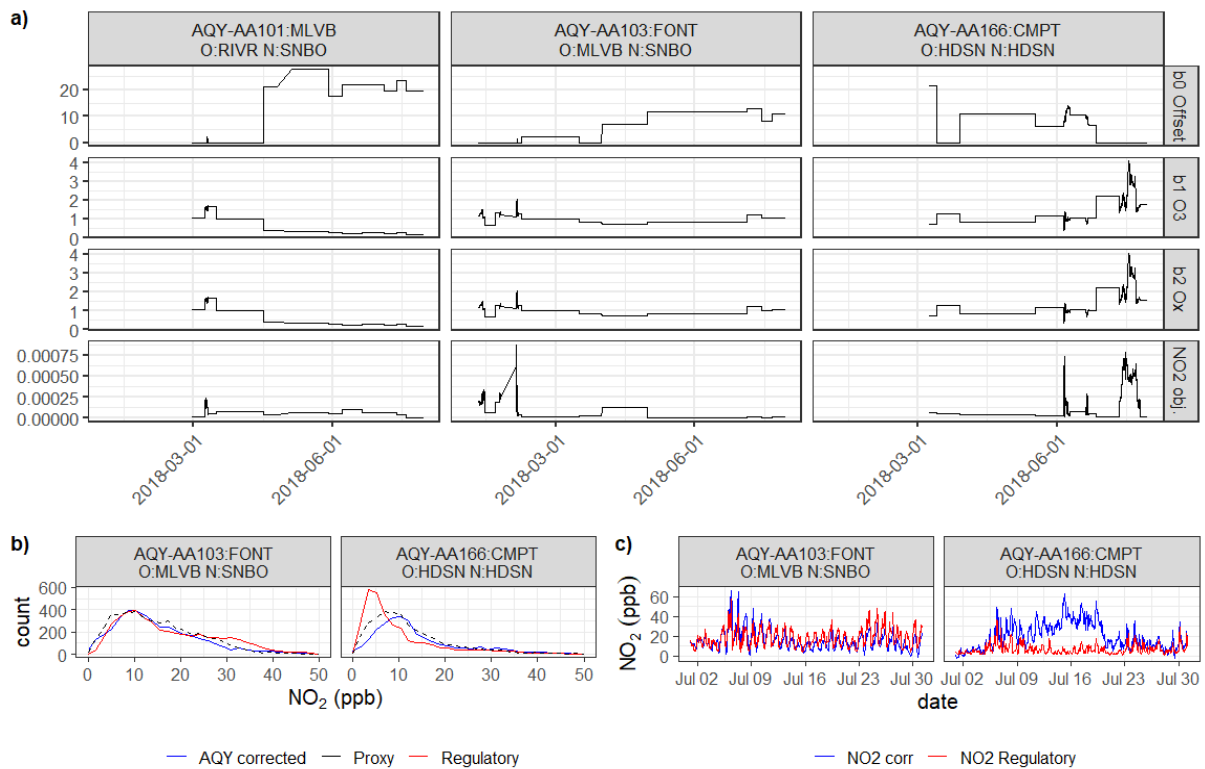

Figure 10. a) Variation over time for three example sites of the fitted parameters: left, MLVB; middle, FONT; right, CMPT; top: offset, $\hat{b}_0$, upper middle: slope parameter $\hat{b}_1$; lower middle: slope parameter $\hat{b}_2$; and bottom: the minimum obtained for the objective function, $D_{KL}$ between sensor data according to equation 2 and the NO$_2$ proxy. At the top of each panel is shown the site designation and the proxies for ozone (O) and NO$_2$ (N). b) Distributions for the month of July of the regulatory station data, the proxy station data and the fitted sensor data (eq 6), for sites at FONT (left) and CMPT (right). c) Time series for July comparing the fitted sensor data, NO$_{2,\text{corr}}$, and the regulatory data at the site with which the sensor was co-located; FONT: left; CMPT: right.

Figure 11 shows hexbin scatter plots of the correlation between the corrected sensor data and the co-located reference station. A hexbin scatter plot for the entire set of corrected sensor data is also presented. The majority of measured NO$_2$ concentrations were low, making the measurement task challenging. The hexbin plots show that the framework correction was generally successful, though clearly less so at the MLVB and SNBO sites. As noted above, the sensor at MLVB failed during April. The variation of the derived parameters for SNBO (figure S5) indicated issues with the proxy, which was confirmed by inspection of the frequency distribution of the NO$_2$ concentrations at the proxy site and at the SNBO regulatory site during June and July, partly explaining the lower success of the management framework for these months. NO$_2$ concentrations can vary considerably at the sub-kilometre scale and the success of the management framework strongly depends on the representativeness of the land use surrounding the reference sites for the low-cost sensor site that is calibrated (Li et al., 2019; van Zoest et al., 2019; Weissert et al., 2019). LAXH is the regulatory site at the Los Angeles airport and its proxy site (CMPT) is in central Los Angeles and may therefore not be a representative site for the local emissions at LAXH. Otherwise, figure 11 shows that the deviations about the 1:1 line were similar at all the sites. The RMSE for the individual sites varied between 4 - 11 ppb (Table S2). For all sensors and sites, the framework-corrected data had RMSE of 7.2 ppb. The higher RMSE, compared to the RMSE using the co-located regulatory $O_3$ and $NO_2$ to correct the data, is mostly related to issues with the proxy (e.g. at CMPT in July, fig. 10b) or to missing data from the proxy site. If proxy data is not available then the method simply uses the latest determined parameters. Specifically, the correction for $e_S$ is not made. Figure S7 shows the error distribution segmented by concentration quartile, for the entire dataset. There was a small concentration-dependent bias and the error distribution was broader for the highest concentration quartile.

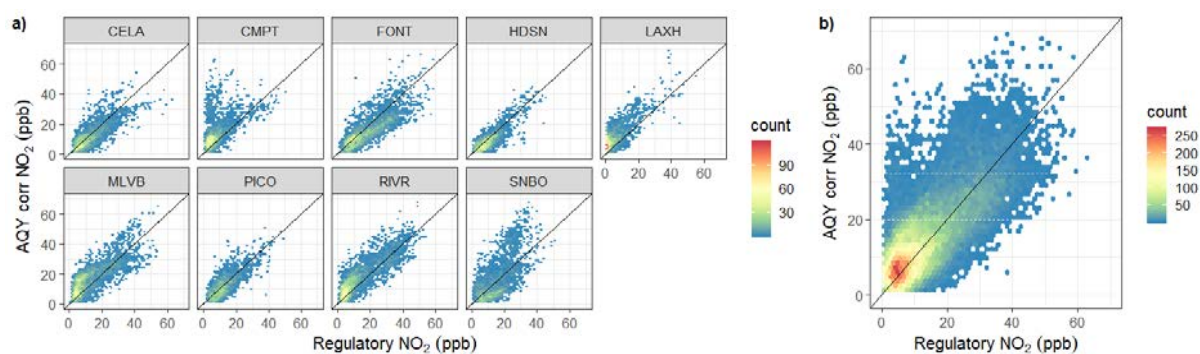

Figure 11 a) Hexbin scatter plots for the framework-corrected data for the individual sites of co-location, for the entire study period. b) Hexbin scatter plot for the set of framework-corrected data (all sensors, all sites).

The potential of low-cost sensors to capture reliably episodes of high concentrations is of great importance for air quality measurements. Figure 12 compares the number of times the low-cost sensor and the regulatory instruments recorded values > $75^{th}$ percentile (20 ppb) per day and indicates that, in general, exceedances will be reliably indicated by the low-cost sensors managed as we have described (Spearman's rank correlation coefficient: 0.81). Comparison with figure 11 shows that the 'false positives' were associated with the site at CMPT, where, as noted above, the proxy comparison failed in July.

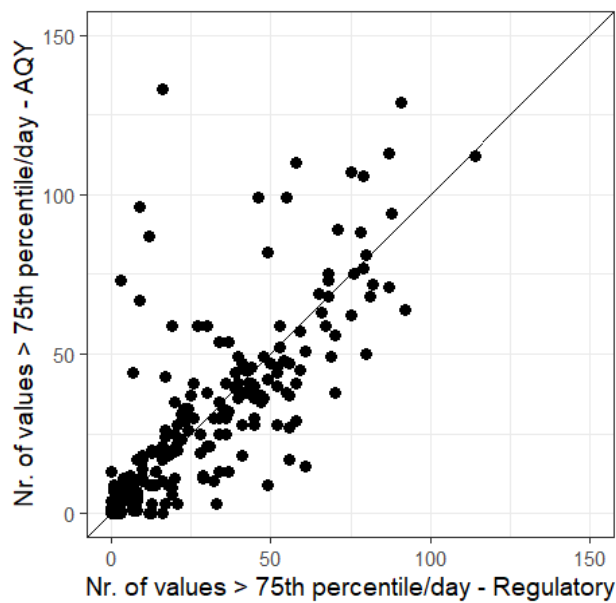

Figure 12. Comparison between number of hourly AQY and regulatory measurements that exceeded the 75$^{th}$ percentile (20 ppb) per day across the whole study period (January to July) for all sites (Spearman's rank correlation coefficient: 0.81).

In our previous paper (part 1: Weissert et al., 2019b submitted) we have shown that the proxy assumptions may not be valid at low wind speed when measured $NO_2$ concentrations are mostly a result of local emissions that are likely different from those at the proxy site. We compared the fit between the corrected sensor $NO_2$ concentrations and the regulatory concentrations for different wind directions and low versus. high wind speed, but did not find any distinct patterns (figures S8 - S11: hexbin scatter plots and error distribution across different wind directions/wind speed). The error distributions across the wind speeds and directions are close to Gaussian with standard deviation not significantly different from the overall RMSE, suggesting that the framework successfully compensated effects related to wind speed or wind direction.

Variation of water vapour pressure is known to have a significant effect on electrochemical sensors – particularly changes of offset, $b_o$ (eq 1) , following rapid changes of humidity (Lewis et al., 2016). In figure S13, we show the distribution of the difference term between the framework-corrected sensor $NO_2$ and the regulatory $NO_2$ across different relative humidity quartiles. No distinct differences can be observed across different relative humidity quartiles, except at the highest, 71 – 100% RH, where the distribution may be bimodal, although there was no significant effect on the correlation with reference data (Fig S11, S12: hexbin scatter plots and error distribution across different RH bands). The error distributions are close to Gaussian with standard deviation not significantly different from the overall RMSE. Thus, the framework and offset error correction compensated for any effect of relative humidity variations.

*3.3 Large local-scale spatial variations in nitrogen dioxide concentration revealed by the low-cost sensor network*

The purpose of the low-cost network has been stated as the supplementary extension of a regulatory network to capture neighborhood-scale variations. The method that we have described uses the regulatory network both to determine and validate the choice of proxy, and then to use the proxy distribution matching to check and re-calibrate if necessary the low-cost sensor network. Indeed, the low-cost sensor network revealed significant $NO_2$ concentration variations that were not captured by the regulatory network, as illustrated in figure 13 and also in figures S14 and S15. Both high and low concentrations of $NO_2$ were very localized and transient, varying between extremes close to the highway network, and also tending to be higher near the mountains at the sides of the valleys. In a subsequent paper, we will show how to use land-use correlations and wind speed-direction information to understand the spatio-temporal variation and identify specific, unusual features, following the ideas given in Weissert et al. (2019a).

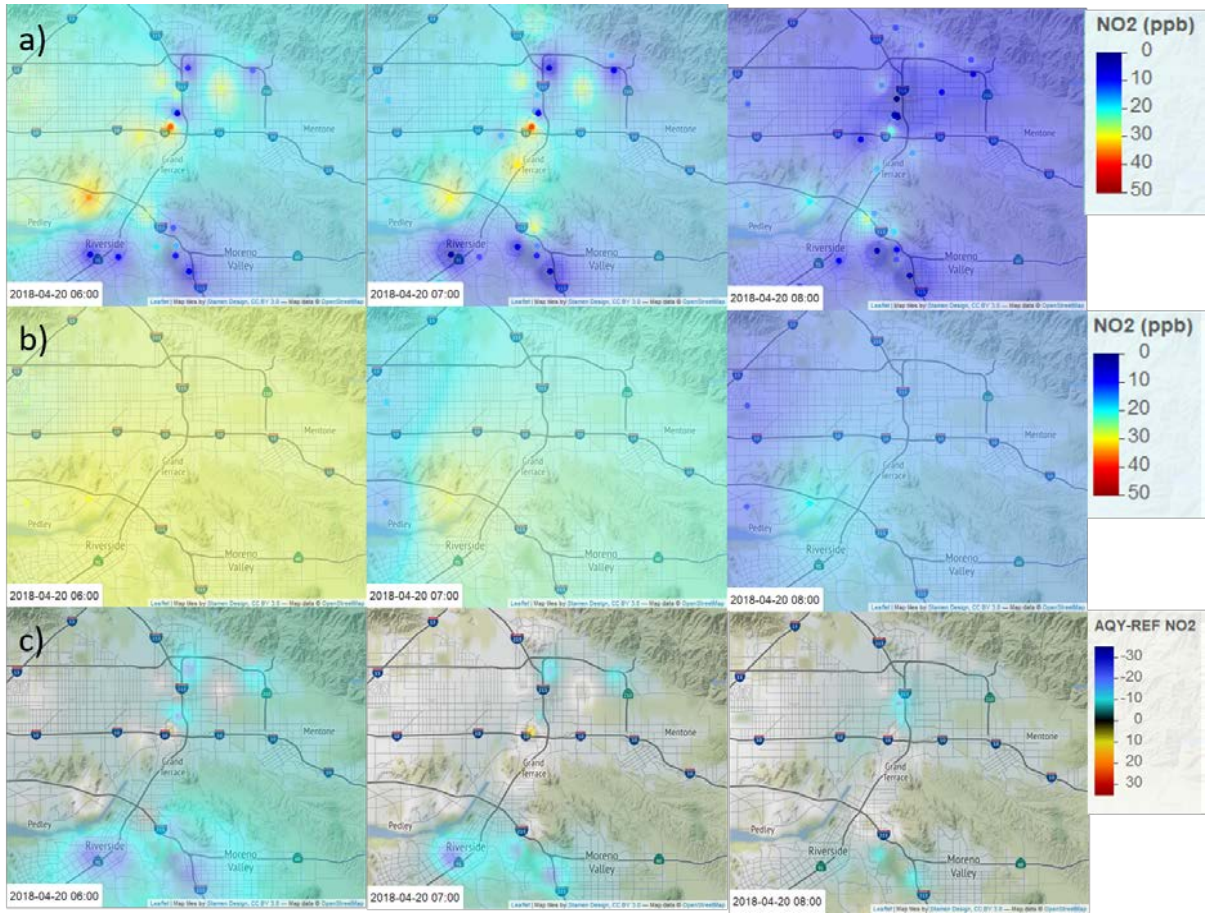

Figure 13. Example of neighbourhood-scale variation in $NO_2$ concentration, in the Riverside-San Bernadino region of Southern California, revealed by the low-cost sensor network, for three successive hours of a particular day. (a): low-cost instrument network (b) reference network only (c) Difference between the low-cost instrument network and the reference network. Interpolation by inverse-distance weighting (power, -2). Symbols and lines mark the major highways.

## 4. Conclusion

In this paper, we have extended a management framework, previously developed to detect and correct for drift in $O_3$ concentrations measured by low-cost air quality sensors, to $NO_2$ measurement by low-cost electrochemical sensors. We used previously selected proxy sites, which have reliable $NO_2$ data, to identify when the sensor data diverged from the expected data. Over a period of time and for appropriately chosen proxies, the sensor and proxy data should be statistically similar. The framework is easily modified to change proxy or to signal uncertainty if conditions occur (such as particular wind direction or speed conditions) where the proxy is known (from other assessment using the reference network) to be unreliable.

When the management framework triggered an alarm, we minimised the Kullback-Leibler divergence between the distribution of the proxy data and the low-cost sensor data by adjusting of the sensor measurement model parameters. Using this approach, we were able to considerably improve the accuracy of the low-cost sensor data as indicated by the lower RMSE. Analysis of the residual errors indicated that the most significant effect was due to uncompensated variation of the baseline current of the electrochemical sensor on a timescale shorter than the distribution averaging timescale. This error was in part spatially correlated and had diurnal variations similar to the variations of ambient temperature, which allowed the error to be partially determined by using the closest proximity reference station with a co-located $NO_2$ sensor as a proxy. Sensor failure could be distinguished through a characteristic time variation of the derived parameters of the sensor measurement model. The results also indicated that failures of this approach, likely due to differences in local emission sources and the lack of suitable proxy sites, could be signalled through consideration of the time variation of the corrected sensor parameters and of the value of the Kullback-Leibler objective function. While the method is robust, it does require a network of reference-grade instruments that is sufficiently diverse to sample all the environments within the zone to be measured. It also requires data availability, not only from the low-cost network but also of ozone and nitrogen dioxide measurements from the reference network.

**Acknowledgements**

This work was funded by the New Zealand Ministry for Business, Innovation and Employment, contract UOAX1413. This work was performed in collaboration with the Air Quality Sensor Performance Evaluation Center (AQ-SPEC) at the South Coast Air Quality Management District (South Coast AQMD). The authors would like to acknowledge the work of Mr. Berj Der Boghossian for his technical assistance with deploying AQY sensor nodes. The authors would like to acknowledge the work of the South Coast AQMD Atmospheric Measurements group of dedicated instrument specialists that operate, maintain, calibrate, and repair air monitoring instrumentation to produce regulatory-grade air monitoring data. DEW acknowledges the support of a fellowship program at the Institute of Advanced Studies, Durham University, UK.

## 6. Competing interests

LW, EM, KA and GSH are employees of Aeroqual Ltd, manufacturer of the sensor nodes used in the study. GSH and DEW are founders and shareholders in Aeroqual Ltd.

## 7. Supplementary Information

Kullback-Leibler divergence: definition; Fitted parameter variation over time for all sites; Spatially-dependent offset error analysis; Framework-and offset error-corrected results for all sites; Correlation of framework- and offset error-corrected sensor data with reference data grouped according to different wind direction, wind speed and humidity; Example maps of variation of mean concentration of $NO_2$ across the region.

Supplementary Information

Hierarchical network design for nitrogen dioxide measurement in urban environments, part 2: sensor calibration

Lena Weissert[1,3,a], Elaine Miles[2,b], Georgia Miskell[1,c], Kyle Alberti[2], Brandon Feenstra[4], Geoff S Henshaw[2], Vasileios Papapostolou[4], Hamesh Patel[2], Andrea Polidori[4], Jennifer A Salmond[3], , David E Williams[1,*].

*Email  david.williams@auckland.ac.nz    ph +64 9 923 9877

1. School of Chemical Sciences and MacDiarmid Institute for Advanced Materials and Nanotechnology, University of Auckland, Private Bag 92019, Auckland 1142, New Zealand

2. Aeroqual Ltd, 460 Rosebank Road, Avondale, Auckland 1026, New Zealand

3. School of Environment, University of Auckland, Private Bag 92019, Auckland 1142, New Zealand

4. South Coast Air Quality Management District, 21865 Copley Drive, Diamond Bar, CA 91765, USA

1. **Kullback-Leibler divergence: definition**

The Kullback-Leibler divergence between two probability distributions *P* and *Q* of a random variable *x* is

$$D_{KL}(P(x)||Q(x)) = \sum_{x} P(x)\ln\left(\frac{P(x)}{Q(x)}\right)$$

2. **Fitted parameter variation over time for all sites**

---

[a] Present address:  Aeroqual Ltd, 460 Rosebank Road, Avondale, Auckland 1026, New Zealand
[b] Present address:  elaine.miles@gmail.com
[c] Present address: Trustpower, 108 Durham St, Tauranga, New Zealand

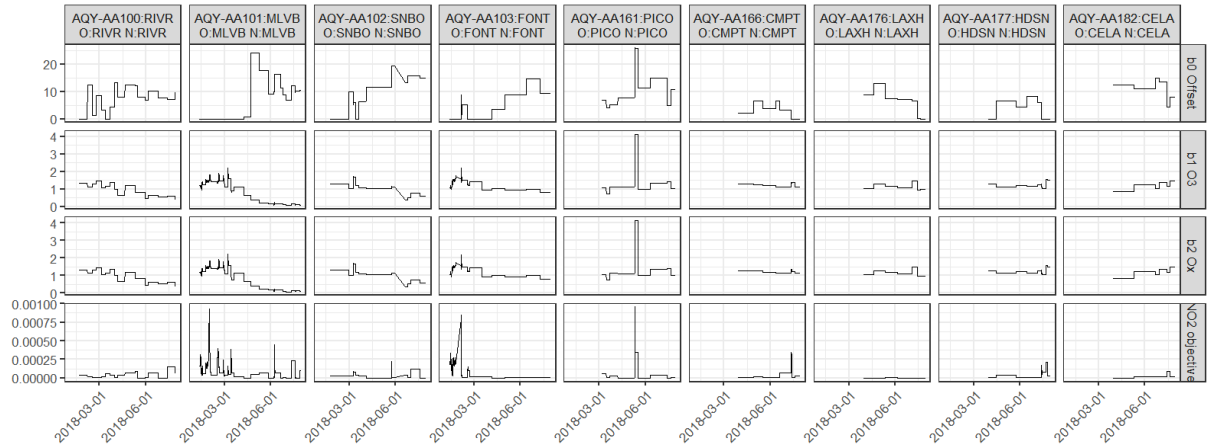

Figure S1. Variation over time of the fitted parameters for all sites, top: offset, $\hat{b}_0$, upper middle: slope parameter $\hat{b}_1$; lower middle: slope parameter $\hat{b}_2$; and bottom: the minimum obtained for the objective function, $D_{KL}$ between sensor data according to equation 2 (main text) and the $NO_2$ proxy. At the top of each panel is shown the site designation and the proxies for ozone (O) and $NO_2$ (N). Co-located data was used here, thus the proxies are the same.

3. **Spatially-dependent offset error analysis**

Table S1 shows the correlation matrix for the difference term between the framework-corrected sensor result and the co-located reference result. The correlation matrix indicates that there is a spatial correlation with a higher correlation coefficient for sites closer to each other (e.g. RIVR and MLVB, SNBO and MLVB, CMPT and PICO). The correlation plots are shown in figure S3

Table S1. Correlation matrix for the difference term between the framework-corrected sensor result and the co-located reference result. Correlations > 0.50 are highlighted in bold.

| Regulatory site | CELA | CMPT | FONT | HDSN | LAXH | MLVB | PICO | RIVR | SNBO |
|---|---|---|---|---|---|---|---|---|---|
| CELA | NA | NA | NA | NA | NA | NA | NA | NA | NA |
| CMPT | **0.53** | NA | NA | NA | NA | NA | NA | NA | NA |
| FONT | 0.28 | 0.37 | NA | NA | NA | NA | NA | NA | NA |
| HDSN | 0.17 | 0.44 | 0.05 | NA | NA | NA | NA | NA | NA |
| LAXH | -0.16 | 0.19 | -0.10 | 0.04 | NA | NA | NA | NA | NA |
| MLVB | 0.26 | 0.40 | **0.84** | 0.30 | -0.17 | NA | NA | NA | NA |
| PICO | **0.58** | **0.70** | -0.06 | 0.39 | 0.30 | 0.37 | NA | NA | NA |
| RIVR | 0.06 | 0.17 | **0.70** | 0.44 | -0.18 | **0.83** | **0.63** | NA | NA |
| SNBO | 0.34 | **0.51** | 0.52 | 0.45 | 0.12 | **0.74** | 0.15 | **0.86** | NA |

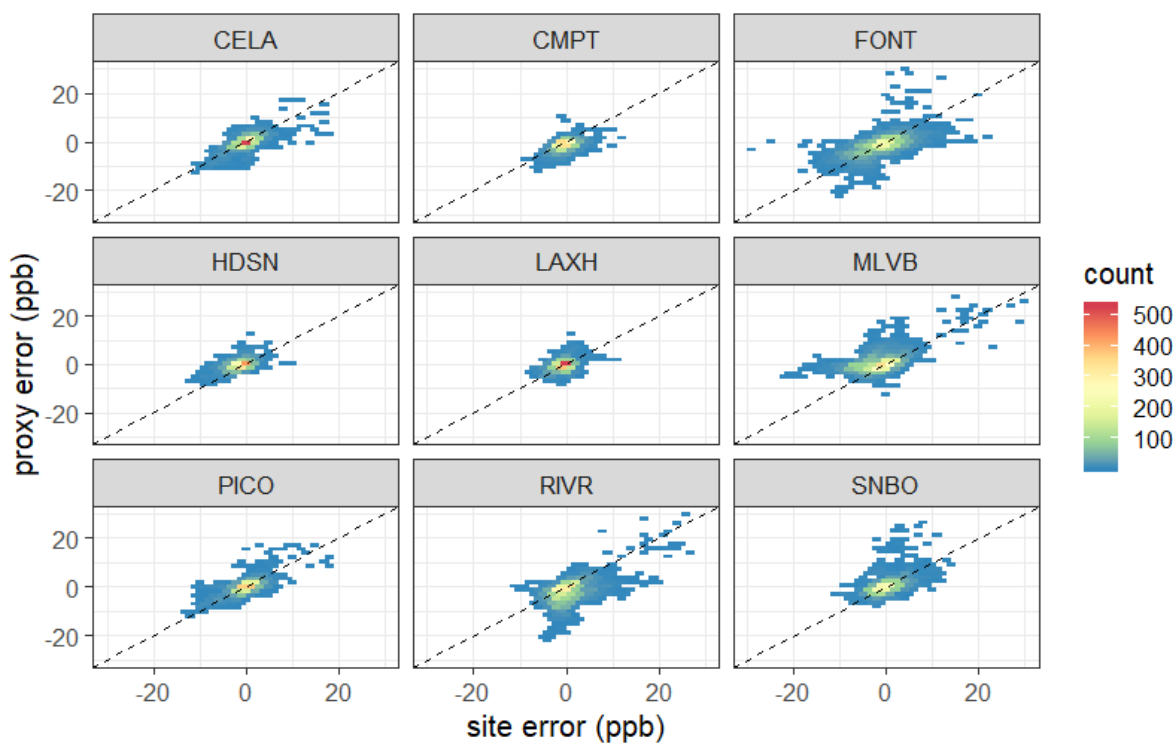

Figure S2. Error correlation: (proxy error: framework-corrected sensor result - co-located reference result at closest proximity proxy) vs (site error: framework-corrected sensor result - the co-located reference result at the reference site).

We used the difference term, $e_s$, from the O$_3$ proxy site (= nearest site) to further correct the framework-corrected sensor result. We used a sigmoid function (eq S1) to damp the error correction and a 3-hour rolling mean to smooth the fluctuations.

$$S(x) = 1/(1 + e^{-k(|x-u|)}) \tag{S1}$$

Here, $x$ is the difference (framework-corrected sensor result minus closest proximity reference result) and $u$ the mean difference term. Empirically, we determined $k = 0.057$ to minimise the resulting RMSE of the sensor result in comparison with the reference station of co-location. Figure S4 shows the error term at each site and the value damped according to eq S1.

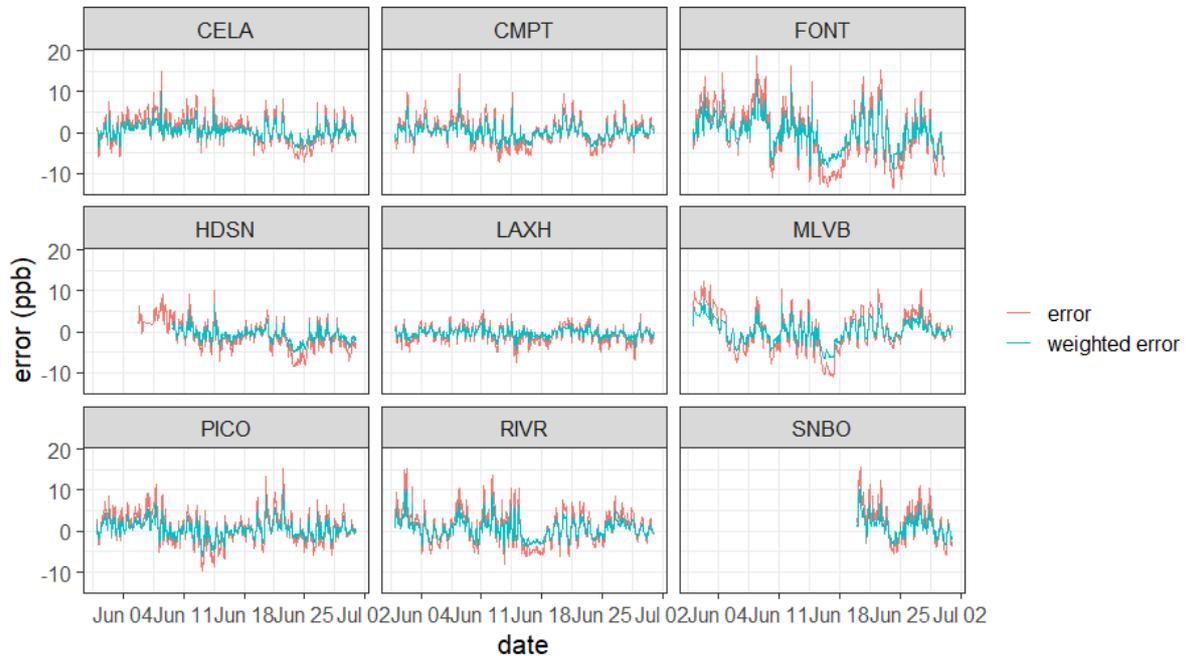

Figure S3. Comparison of the error term at each site and that damped according to eq S1.

4. **Framework-corrected results for all sites**

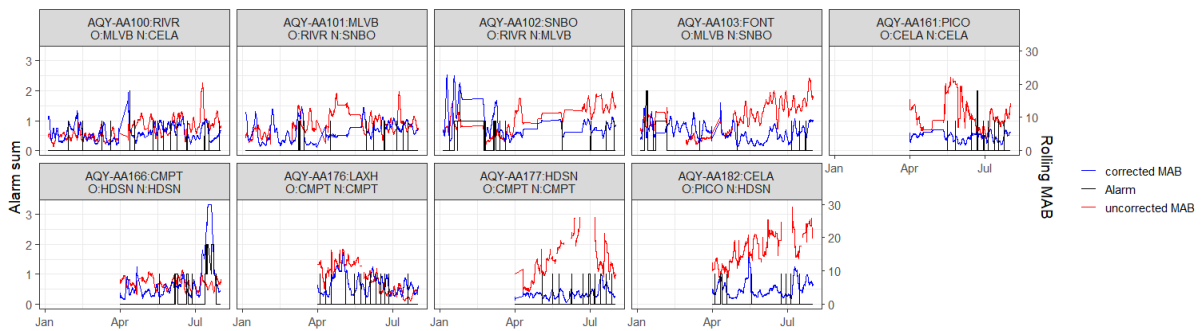

Figure S4. The number of alarm signals generated by the low-cost sensor data in comparison with the proxy data (left-axis), and uncorrected vs. corrected mean absolute bias (MAB) / ppb running over 72 hr of the low-cost sensor data with respect to the co-located regulatory station (right-axis), when the sensor data are corrected using proxies, including the spatially-correlated

error term determined using the closest proximity other site as proxy. At the top of each panel is shown the site designation and the proxies for ozone (O) and NO$_2$ (N).

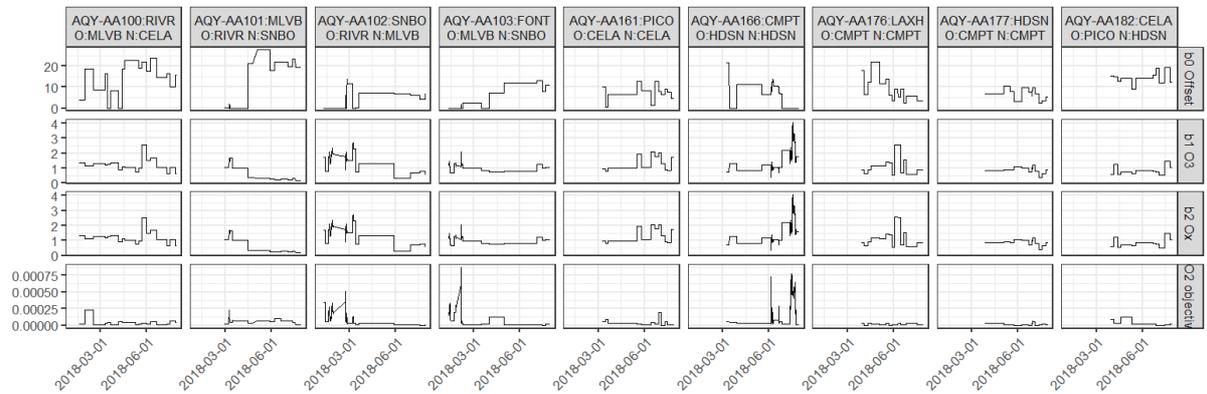

Figure S5. Variation over time for all sites of the fitted parameters: offset, $\hat{b}_0$, slope parameter $\hat{b}_1$, slope parameter $\hat{b}_2$, and bottom: the minimum obtained for the objective function, $D_{KL}$ between sensor data according to equation 2 and the NO$_2$ proxy. The sensor data are corrected using proxies. At the top of each panel is shown the site designation and the proxies for ozone (O) and NO$_2$ (N).

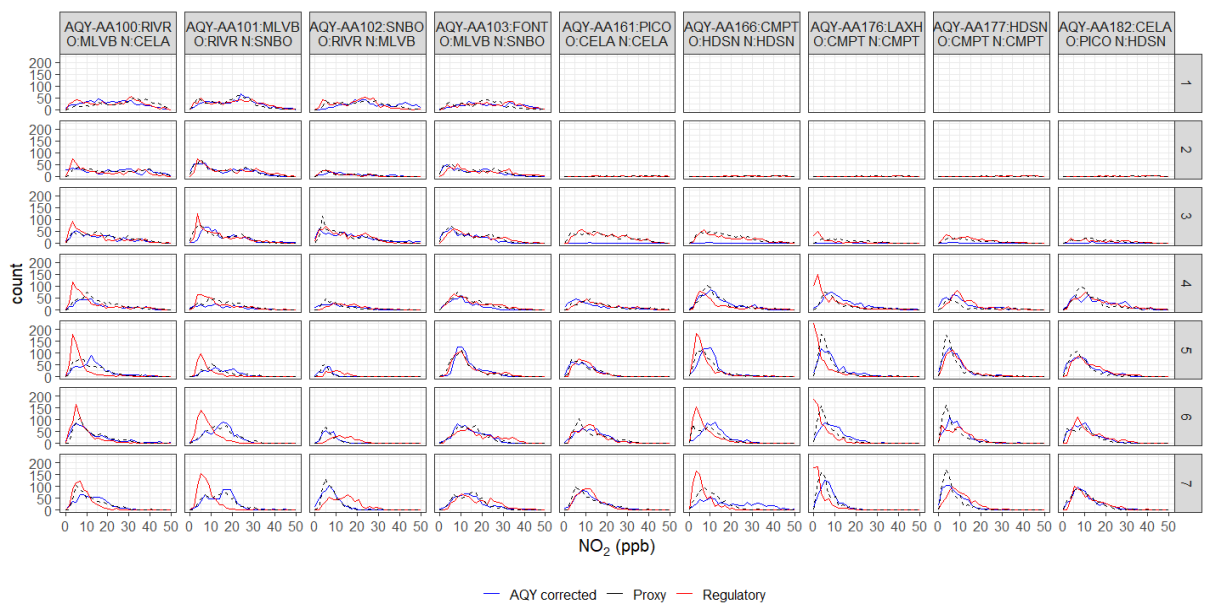

Figure S6. Distributions for different months of the regulatory station data, the proxy station data and the fitted sensor data.

Table S2. Summary statistics comparing the uncorrected and corrected AQY $NO_2$ data against the regulatory $NO_2$ data, for uncorrected, framework-corrected data using co-located reference data and sensor data that are framework-corrected with $e_s$ applied using proxy data.

| Regulatory Site | Uncorrected | | | Framework-corrected (co-located) + $es$ | | | Framework-corrected + $es$ (proxies) | | |
| --- | --- | --- | --- | --- | --- | --- | --- | --- | --- |
| | $R^2$ | MAB | RMSE | $R^2$ | MAB | RMSE | $R^2$ | MAB | RMSE |
| RIVR | 0.72 | 6.34 | 8.13 | 0.81 | 3.78 | 5.32 | 0.74 | 5.35 | 6.81 |
| MLVB | 0.53 | 7.86 | 9.72 | 0.76 | 4.90 | 6.75 | 0.63 | 5.90 | 7.30 |
| SNBO | 0.26 | 9.12 | 11.18 | 0.73 | 3.59 | 4.89 | 0.48 | 7.89 | 10.14 |
| FONT | 0.47 | 10.13 | 11.92 | 0.72 | 4.64 | 6.30 | 0.67 | 5.42 | 6.96 |
| PICO | 0.07 | 9.89 | 11.99 | 0.69 | 2.69 | 4.69 | 0.61 | 4.02 | 5.15 |
| CMPT | 0.70 | 6.36 | 7.45 | 0.83 | 2.32 | 3.08 | 0.25 | 6.94 | 10.61 |
| LAXH | 0.60 | 4.90 | 6.62 | 0.85 | 2.22 | 2.95 | 0.60 | 5.92 | 7.35 |
| HDSN | 0.46 | 10.06 | 11.82 | 0.82 | 2.67 | 3.58 | 0.72 | 3.19 | 4.19 |
| CELA | 0.60 | 14.71 | 15.88 | 0.79 | 2.53 | 4.09 | 0.55 | 4.58 | 6.55 |

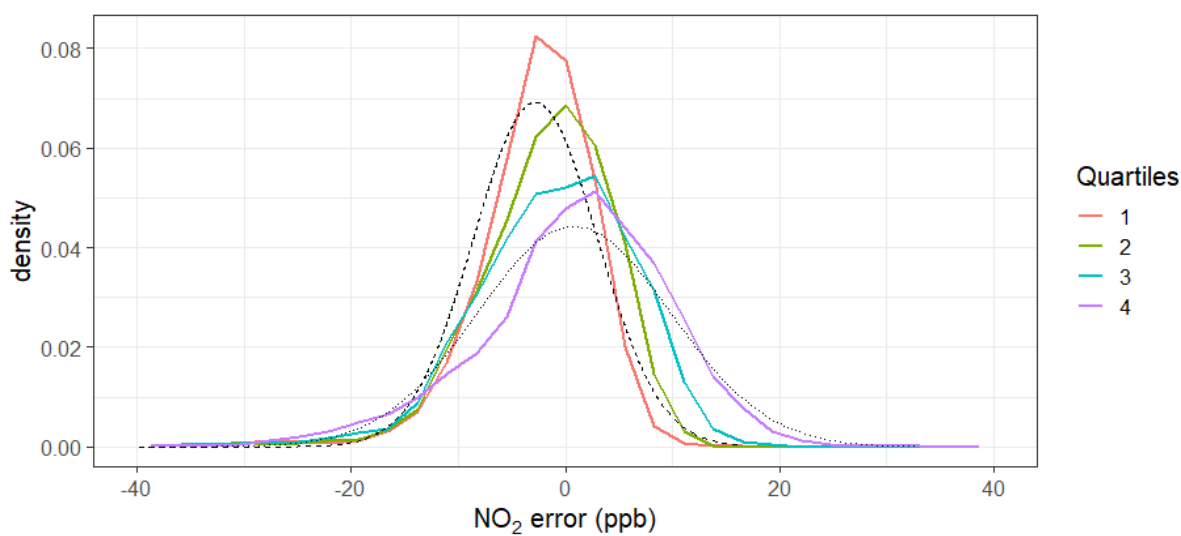

Figure S7. Distribution of differences between framework- and $e_S$ – corrected sensor data and regulatory $NO_2$ concentrations across different quartiles (Quartiles: 5, 10, 18, 82 ppb). The dashed line is a Gaussian distribution for Quartile 1, the dotted line is the Gaussian distribution for Quartile 4, where the Gaussian has the same mean and standard deviation as the data.

5. **Correlation of proxy-corrected sensor data with reference data grouped according to different wind direction, wind speed and humidity**

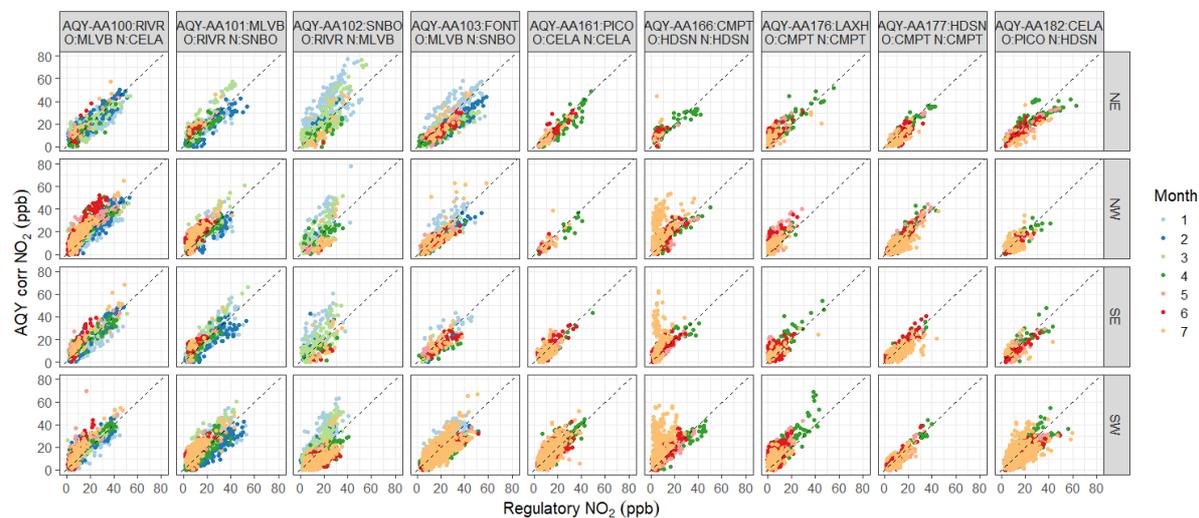

Figure S8. Hex-bin scatterplots showing the framework-corrected sensor data against the regulatory data grouped into different wind directions. The dashed line is the 1:1 line.

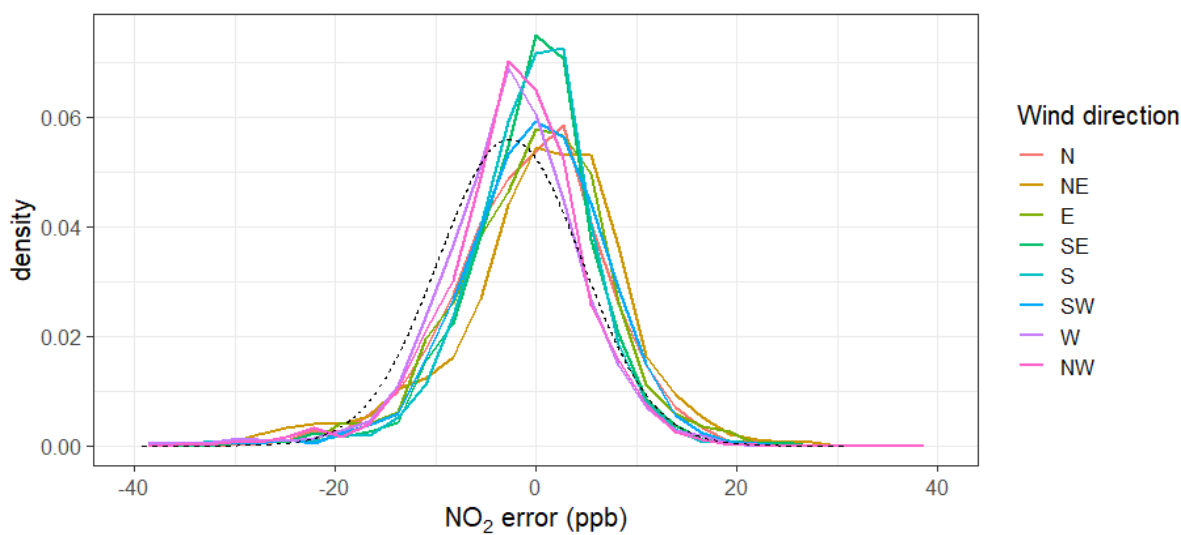

Figure S9. Error (regulatory $NO_2$ – error corrected AQY $NO_2$) distribution, segmented by wind direction. The dashed line is a Gaussian distribution for winds from the west, with the same mean and standard deviation as the data.

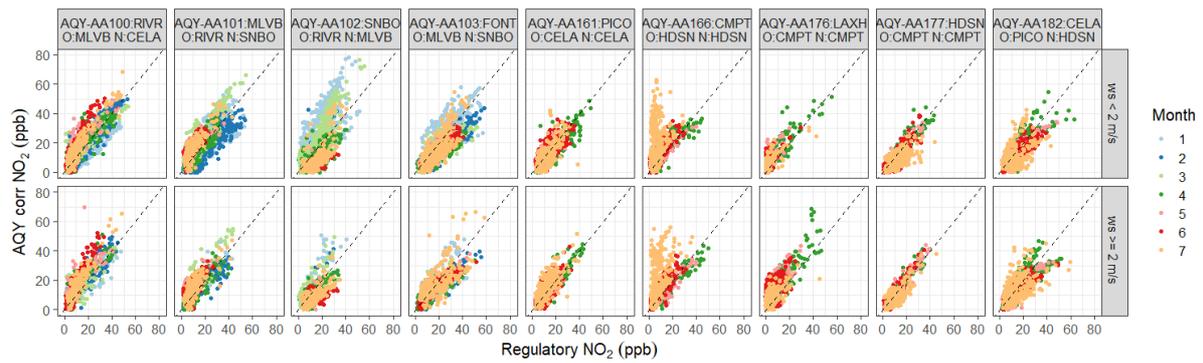

Figure S10. Hexbin scatterplots showing the framework-corrected sensor data against the regulatory data when wind speed was low (< 2 m s$^{-1}$) and high (> 2 m s$^{-1}$). The dashed line is the 1:1 line.

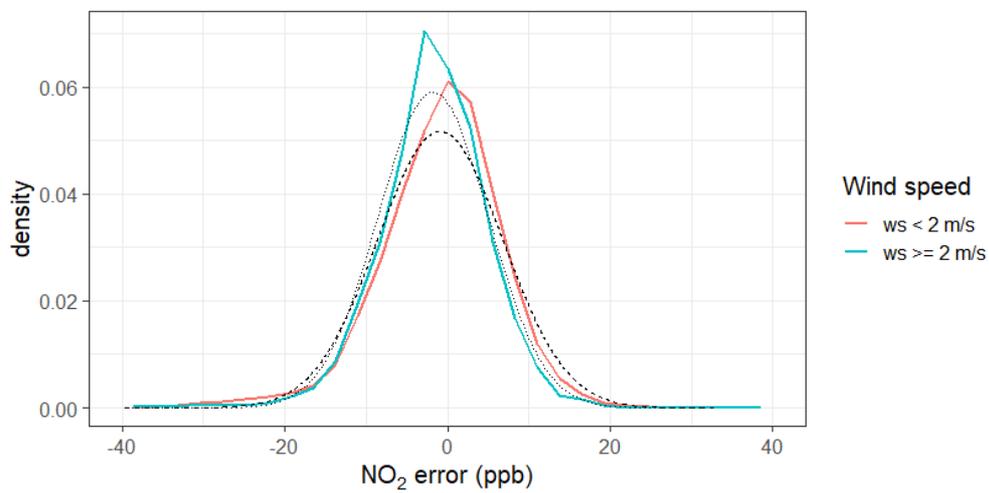

Figure S11. Error (regulatory NO$_2$ – error corrected AQY NO$_2$) distribution, segmented by wind speed. The dashed line is a Gaussian distribution for ws < 2m/s, the dotted line is the Gaussian distribution for ws > 2m/s, where the Gaussians have the same mean and standard deviation as the data

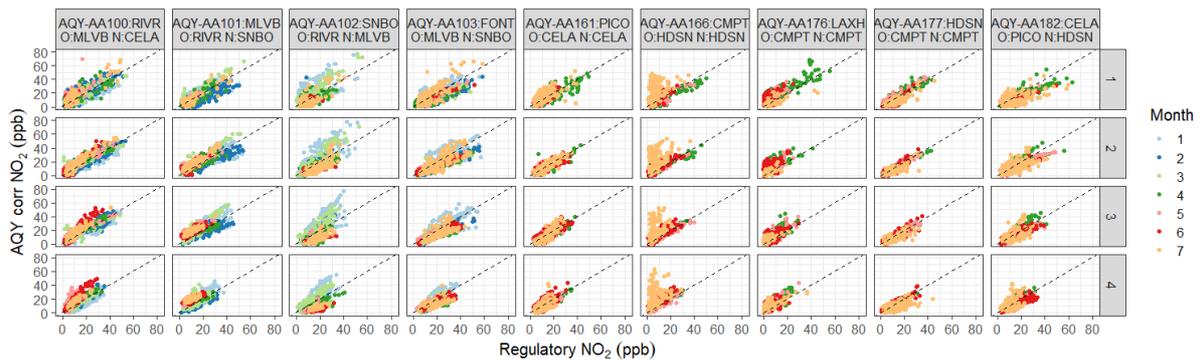

Figure S12. Hex-bin scatterplots showing the framework-corrected sensor data against the regulatory data grouped by relative humidity quartiles (Quartiles: 1: 37.8, 2: 57.5, 3: 70.7, 4: 99.9%). The dashed line is the 1:1 line.

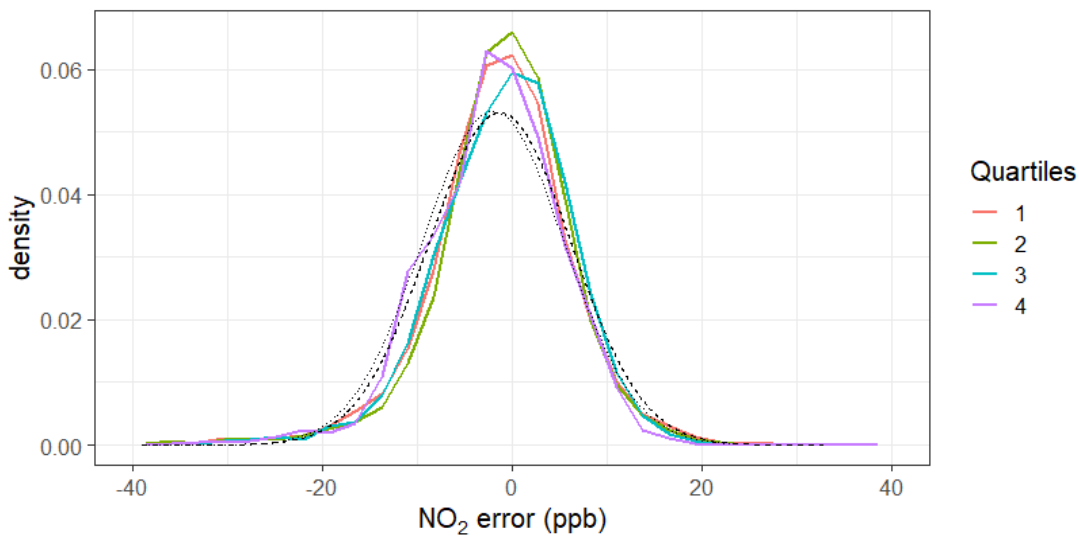

Figure S13. Frequency distribution of the error (regulatory $NO_2$ – error corrected AQY $NO_2$) across different relative humidity quartiles (Quartiles: 1: 37.8, 2: 57.5, 3: 70.7, 4: 99.9%). The dashed line is a Gaussian distribution for Quartile 1, the dotted line is the Gaussian distribution for Quartile 4, where the Gaussians have the same mean and standard deviation as the data.

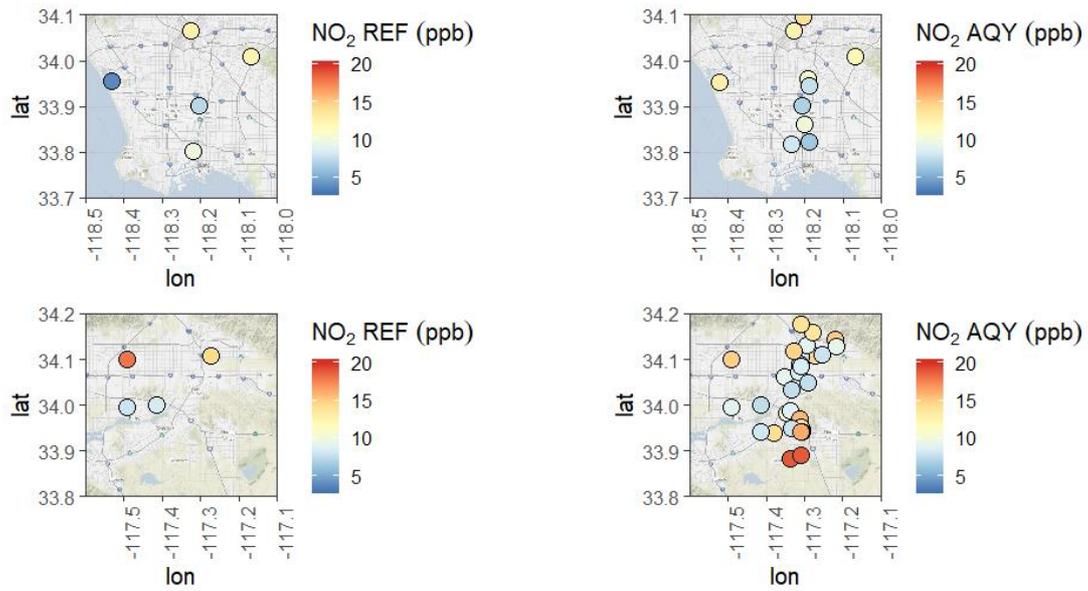

figure S14. Mean NO$_2$ concentration for June 2018, comparing measurements at the reference sites and at the sensor sites for the two regions, Los Angeles City (LA) and Inland Empire (IE: San Bernardino – Riverside).

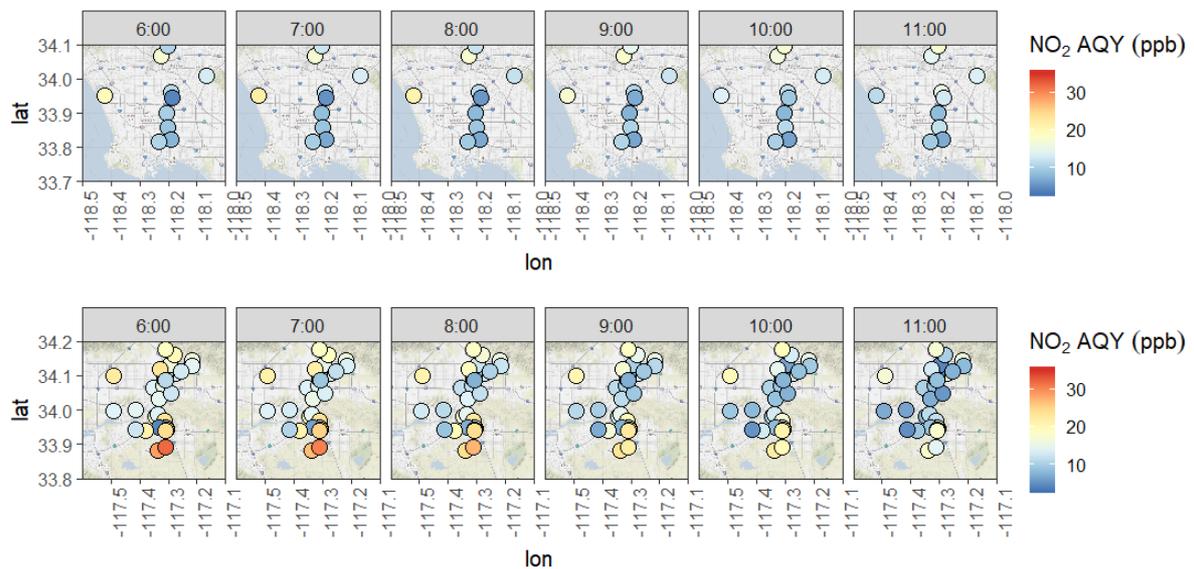

Figure S15. Diurnal mean NO$_2$ concentration for June 2018, at the sensor sites for the two regions, Los Angeles City (LA) and Inland Empire (IE: San Bernadino – Riverhead).